\documentstyle[12pt]{article}

\renewcommand{\a}{\alpha}
\renewcommand{\b}{\beta}
\renewcommand{\c}{\gamma}

\newcommand{\k}{\kappa}

\newcommand{\shalf}{\frac{1}{2}}
\newcommand{\pa}{\partial}
\begin{document}

\topmargin 0pt
\oddsidemargin 5mm

\newcommand{\NP}[1]{Nucl.\ Phys.\ {\bf #1}}
\newcommand{\AP}[1]{Ann.\ Phys.\ {\bf #1}}
\newcommand{\PL}[1]{Phys.\ Lett.\ {\bf #1}}
\newcommand{\NC}[1]{Nuovo Cimento {\bf #1}}
\newcommand{\JMP}[1]{Jour.\ Math.\ Phys.\ {\bf #1}}
\newcommand{\PR}[1]{Phys.\ Rev.\ {\bf #1}}
\newcommand{\PRL}[1]{Phys.\ Rev.\ Lett.\ {\bf #1}}
\newcommand{\JL}[1]{JETP.Lett.\ {\bf #1}}
\newcommand{\JETP}[1]{\ Jour.\ Eksp.\ Teor.\ Phys.\ {\bf #1}}
\newcommand{\YP}[1]{Yad.\ Phys.\ {\bf #1}}
\renewcommand{\thefootnote}{\fnsymbol{footnote}}

\begin{titlepage}
\setcounter{page}{0}
\rightline{Preprint YERPHY-1398(9)-93}

\vspace{2cm}
\begin{center}
{\Large Pseudoclassical  Foldy-Wouthuysen transformation and
 the canonical quantization of the D=2n dimensional
 spinning particle in the external electromagnetic field}
\vspace{1cm}

{\large Grigoryan G.V., Grigoryan R.P.} \\
\vspace{1cm}
{\em Yerevan Physics Institute, Republic of Armenia}\\
\end{center}

\vspace{5mm}
\centerline{{\bf{Abstract}}}

The canonical quantization of $D=2n$ dimensional Dirac
spinning particle in the external electromagnetic field is
carried out in the gauge which allows to describe
simultaneously particles and antiparticles (massive and
 massless) already at the classical level.  Pseudoclassical
Foldy-Wouthuysen transformation is used to obtain canonical
(Newton-Wigner) coordinates and in terms of this variables
the theory is quantized.  The connection of this
 quantization with the Blount picture of Dirac particle in
the external electromagnetic field is discussed.

\vfill
\centerline{\large Yerevan Physics Institute}
\centerline{\large Yerevan 1993}

\end{titlepage}
\newpage
\renewcommand{\thefootnote}{\arabic{footnote}}
\setcounter{footnote}{0}

\section{Introduction}
\indent

In papers [1-3]
 the relativistic spinning particle was
canonically  quantized  in  the  free  case  (in  $D=2n$
dimensions) and in the external electromagnetic field
$(D=4)$.  The pseudoclassical description of the
 particle (see [4-7])
 was used in  these papers  and  the
quantization   scheme   was   characterized   by
introduction  at  the  classical  level  of   all   gauge
fixing constraints. This quantization was shown to result
in  the  Dirac theory in the Foldy-Wouthuysen representation
in the free particle case (see also \cite{GTY1} ), while  for
the   spinning  particle  in  the external electromagnetic
field  a generalization of  the  Blount's picture \cite{BL}
 was obtained. The quantization was carried out not  in the
terms of the initial variables of the theory, but in the
terms of new (Newton-Wigner )  variables,  for  which  the
quantum commutation relations are canonical. The transition
to  canonical variables  prior   to   quantization   seems
more   appropriate, considering that in the case of the
relativistic particle  in  the external electromagnetic
field the  operator  realization  of  the theory in terms of
the initial variables appears impossible due to the
    complexity of corresponding Dirac brackets.

In this paper a method of constructing  of  the  Newton-Wigner
type variables at the classical level  using  the
pseudoclassical analog of the Foldy-Wouthuysen
transformation is  proposed.   This allows  to  obtain
straightforwardly  the  relation  between  the canonical and
initial variables bypassing the calculation  of  the Dirac
brackets  with  the  subsequent  diagonalization  of  these
brackets. The investigation  is  carried  out  in  the
space-time dimensions $D=2n$.  Note  that   pseudoclassical
Foldy-Wouthuysen transformation was used in \cite{BCL2} to obtain
the expression  for  the hamiltonian of the electron
interacting with the external constant magnetic field, which
after quantization automatically  brought  to diagonalized
hamiltonian.

\section{Constraints}
\indent

    Consider the action of the theory, describing the
relativistic spinning particle in the external
electromagnetic field  in   D=2n \cite{BM},\cite{BCL},
 \cite{GT}
\begin{eqnarray}
&& L={\shalf}\int d\tau\bigl[\frac{\left(\dot x^\mu\right)^2}
{e}+em^2-i\left(\xi_\mu\dot \xi^\mu-\xi_{D+1}
\dot \xi_{D+1}\right)
-i\chi\left(\frac{\xi_\mu\dot x^\mu}{e}
-m \xi_{D+1}\right) + \nonumber\\
&&\hspace {4cm}+ 2g\dot x^\mu A_\mu+
igeF_{\mu\nu}\xi^\mu \xi^\nu \bigr],
\end{eqnarray}
here $ x^\mu$  is particle coordinate,
 $\mu= 0, 1, 2, \ldots, D-1$, $\xi^\mu$ is
   Grassmann variables, describing
 spin degrees of  freedom, $\xi_{D+1}, \chi$ and $e$  are
additional fields ($e$ is an even element, $\xi_{D+1}$ and
 $\chi$ are odd elements  of Grassmann algebra) , $g$ is the
charge of  the  particle, $A^\mu$   is  the vector-potential
of the electromagnetic field, $F_{\mu\nu} = \pa_\mu A_\nu -
\pa_\nu A_\mu$ ,  the overdote denotes the differentiation
over $\tau$ along the  trajectory; the derivatives over
 Grassmann variables are left.  Following the steps analogous
 to those of \cite{GG3}  we come  to
   the complete set of constraints
\begin{equation}
\label{ConstraintA}
\Phi_\mu =
\pi_\mu-\frac{i}{2}\xi_\mu, \mu=0, 1, 2, \ldots, D-1;\quad \Phi_D =
\pi_{D+1}+\frac{i}{2}\xi_{D+1},
\end{equation}
\begin{equation}
\label{ConstraintB} \Phi_{D+3} = {\cal P}^2-m^2
-igF_{\mu\nu}\xi^\mu\xi^\nu,\quad \Phi_{D+4}=
x_O^\prime,\quad\Phi_{D+5}=\pi_e,\quad\Phi_{D+6}=
e+\frac{1} {\tilde \omega},
\end{equation}
\begin{equation}
\label{ConstraintC}
\Phi_{D+1}={\cal P}_\mu\xi^\mu-m\xi_{D+1},\quad
\Phi_{D+2}=a\xi_O+b\xi_{D+1},
\end{equation}
here $\pi_\mu, \pi_{D+1}, \pi_e$
 are momenta, canonically conjugate to
$\xi^\mu, \xi_{D+1}$ and $e$ - respectively, ${\cal P}_\mu=
P_\mu-gA_\mu, {\cal P}_O=-\k \tilde \omega,
\tilde \omega =\sqrt{{{\cal P}_i}^2+m^2+
igF_{\mu\nu}\xi^\mu\xi^\nu}$. Recall [3], that
$x^\prime_O=x_O-\k\tau$ and the constraint
$\Phi_{D+4}=x_O-\k\tau$  , which  is one of the gauge
fixing constraints, was transformed into
Eq.(\ref{ConstraintB}) by a canonical transformation of
variables $x^\mu, P_\mu$ to $x^{\prime\mu}, P_\mu^\prime$
 defined by the relations
\begin{equation}
x_O^\prime =
x_O-\k\tau,\quad x^{\prime i}=x^i,\quad P_\mu^\prime=P_\mu
\end{equation}
(the corresponding generating function is $W =
   x^\mu P^\prime_\mu-\tau\k P^\prime_O$ ).  The value
 $\k=+1$ corresponds  to  a  particle, $\k=-1$
corresponds  to  a antiparticle. For the independent
variables of the theory we choose $x^i, {\cal P}_i, \xi^i$.

\section{Foldy-Wouthuysen Transformation}
\indent

Let us  now consider  a   pseudoclassical
canonical Foldy - Wouthuysen transformation with a generator
of  the  infinitesimal  canonical transformation \cite{BCL2}
\begin{equation}
\label{Generator}
S_{\rm cl}=-2i\left({\cal P}_j
\xi_j\right)\xi_{D+1}\theta,
\end{equation}
where $\theta$
is  a  function  of  the  variables   of
  the   theory
which  will  be  specified  later.  The  result  of   the
finite canonical  transformation  of   any   dynamical
quantity $f$   is given by the expression \cite{SM}
\begin{equation}
\label{CanonTrans}
\tilde f
=\widetilde{{e^{S_{\rm {cl}}}f}} = f+{\{f,S_{\rm cl}\}}^*+
\frac{1}{2!}{\{\{f,S_{\rm cl}\},S_{\rm cl}\}}^*+\ldots,
\end{equation}
where
$\{\dot,\dot\}^*$ denotes the Dirac brackets for the
(\ref{ConstraintA}).For the variables of the theory
 we have
\begin{equation}
\label{DiracBrackets}
{\{x_\mu,{\cal P}_\nu\}}^* = g_{\mu\nu}, \quad
\{{\xi_\mu, \xi_\nu\}}^* = ig_{\mu\nu}, \quad
{\{\xi_{D+1}, \xi_{D+1}\}}^* = -i, \quad {\{{\cal P}_\mu, \cal
 P_\nu\}}^* = gF_{\mu\nu}
\end{equation}
(all other  brackets vanish).

    Applying Eq. (\ref{CanonTrans}) to a function  A  of  the
 independent  variables $x^i,{\cal P}_i, \xi^i$   and taking into
account the relations
\begin{eqnarray}
&& {\{A\xi_{D+1},S_{\rm {cl}}\}}^*=
 A(2\theta) ({\cal P}_j\xi_j),   \nonumber \\
&&  {\{A({\cal P}_j\xi_j),S_{\rm {cl}}\}}^*=
 -(2\theta)\gamma A\xi_{D+1}+({\cal P}_j\xi_j)\xi_{D+1}R_1, \\
&& {\{A({\cal P}_j\xi_j)\xi_{D+1},S_{\rm {cl}}\}}^*=0, \nonumber
\end{eqnarray}
where $\gamma=i{\{({\cal P}_i\xi_i),({\cal P}_j\xi_j)\}}^*
 = {{\cal P}_i}^2+igF_{ij}\xi_i\xi_j$ , $\rm R_1$
 is a  function  of  the   variables of
the theory, we find for $\tilde A$ the expression
\begin{eqnarray}
&& \tilde A = A -
\frac{i}{\sqrt\c}}{\{A,({\cal P}_j\xi_j)\}^*
\xi_{D+1}sin(2\theta \sqrt\c)+  \nonumber \\
&& + {\frac{i}{\c}}
{\{A,({\cal P}_j\xi_j)\}}^*({\cal P}_k\xi_k)
 \left(\cos(2\theta\sqrt \c)-1\right) + \left({\cal
P}_i\xi_i\right) \xi_{D+1}\rm R_2,
\end{eqnarray}
where $R_2$, like $R_1$, depends on the  variables  of   the
theory, the explicit expressions  of $R_2$   and $\rm R_1$
are immaterial. If  we now specify the function $\theta$ by
 taking $tg(2\theta\sqrt\c)=\frac{\sqrt\c}{m}$, and hence
$\sin(2\theta\sqrt\c)=\frac{\sqrt\c}{\omega}, \quad
\cos(2\theta\sqrt\c)=\frac{m}{\omega}$, then
\begin{equation}
\label{TransVar}
\tilde A = A - i{\{A,({\cal P}_i\xi_i)\}}^*
\frac{\xi_{D+1}(\omega+m) + ({\cal P}_j\xi_j)}{\omega(\omega+m)}
+ ({\cal P}_i\xi_i)\xi_{D+1}{\rm R_2},
\end{equation}
 where $\omega =
\sqrt {{\cal P}_i^2+ m^2+igF_{ij}\xi_i\xi_j}=\sqrt {\c+m^2} $.

\section{Final Dirac Brackets}
\indent

The next  step will  consist of  the prove, that   the
 variables $\tilde x^i, \tilde {\cal P}_i, \tilde \xi^i$
obtained  by  application of Eq.(\ref{TransVar}) to the
variables $x^i, {\cal P}_i, \xi^i$ are Newton-Wigner
variables,  i.e.  the final  Dirac brackets
for    these    variables are  canonical.  To
achieve    this   end    we    use   the     relation
\begin{equation}
\label{BracketProperty}
{\{\tilde A,\tilde B\}}_{D\left(\Phi\right)}=
{\{A^\prime,B^\prime\}}_{D\left(\Phi\right)},
\end{equation}
where $A^\prime \equiv \tilde A \vert_{\Phi = 0
\equiv \tilde B\vert_{\Phi = 0}$,
${\{\ldots,\ldots\}}_{D(\Phi)}$ denotes  the Dirac
 brackets for the      complete set of  constraints
(\ref{ConstraintA})-(\ref{ConstraintC})
.  The equation  Eq.(\ref{BracketProperty}) is a
 reflection  of the   property   of   the  Dirac brackets
which  states that   the Dirac brackets of   the
  constraints with any  dynamical quantity vanish.
    From  the constraints Eq. (\ref{ConstraintC})  we
find
\begin{equation}
\label{ExtraXi}
 \xi_{D+1} =
-\frac{a\left({\cal P}_j\xi_j\right)}{\tilde\beta}, \quad
\xi_O=\frac{b\left({\cal P}_j\xi_j\right)}{\tilde\beta},
\end{equation}
where  $\tilde\beta = -b\k\omega+am$.
Substituting   now    Eq. (\ref{ExtraXi})   in Eq.
(\ref{TransVar})   we    get
\begin{equation}
\label{PrimedVar}
A^\prime \equiv \tilde A \vert_{\Phi=0} =
A+i{\{A,({\cal P}_i\xi_i)\}}^* ({\cal P}_j\xi_j)\frac{(b\k
 + a)} {\tilde \beta(\omega+m)}.
\end{equation}

Using   Eq.
(\ref{PrimedVar})   we   find   for   variables $x^{\prime
i}, {\cal P}_i^\prime, \xi^{\prime i}$ the  expressions
\begin{eqnarray}
\label{PrimedIndVar}
&&x^\prime_i  = x_i-i\xi_i ({\cal P}_j\xi_j)
 \frac{(b\k + a)}{\tilde\b
 (\omega + m)} \equiv q_i,  \nonumber\\
&&{\cal P}_i^\prime={\cal P}_i +
igF_{ij}\xi_j ({\cal P}_k\xi_k)\frac{(b\k+a)}
 {\tilde \b(\omega + m)}\equiv\pi_i,  \\
&&\xi_i^\prime = \xi_i + {\cal P}_i
({\cal P}_j\xi_j)\frac{(b\k+a)}
{\tilde \b(\omega+m)} \equiv \psi_i.  \nonumber
\end{eqnarray}

Consider now the   right   side   of  the equation
(\ref{BracketProperty}).  We have  by   definition
\begin{equation}
\label{FinalBracket}
{\{A^\prime,B^\prime\}}_{D(\Phi)}=
{\{A^\prime,B^\prime\}}^{**}
-{\{A^\prime,\varphi_r\}}^{**}C^{-1}_{rr^\prime}
{\{\varphi_{r^\prime},B^\prime\}}^{**}.
\end{equation}
Here $\varphi_r=(\Phi_{D+1},\Phi_{D+2}),
{\{\ldots,\ldots\}}^{**}$   stands for the Dirac brackets
 for   a subset   of constraints
 (\ref{ConstraintA}),(\ref{ConstraintB}), $C^{-1}$  is
   the  inverse matrix of
\begin{equation}
C_{rr^\prime}={\{\varphi_r,\varphi_{r^\prime}\}}^{**}
\end{equation}
Now we'll   take   an   advantage of  the
special structure  of  the constraints (\ref{ConstraintB}):
one  of each  pair of constraints is   a    canonical
variable  .  This  allows  to prove   that
\begin{equation}
\label{PrelimDiracBracket}
{\{F,G\}}^{**}={\{F,G\}}^{*}
\end{equation}
for any     dynamical    variables
  $F$ and $G$ (see   e.g.\cite{GTY2}).
 With account    of Eq. (\ref{PrelimDiracBracket} ) the
 formula  (\ref{FinalBracket} )    takes    the   form
\begin{equation}
\label{FinalDiracBracket}
{\{A^\prime,B^\prime\}}_{D(\Phi)}=
{\{A^\prime,B^\prime\}}^*
-{\{A^\prime,\varphi_r\}}^* C^{-1}_{rr^\prime}
{\{\varphi_{r^\prime},B^\prime\}}^*.
\end{equation}
The  matrix    $C$ is given by
\begin{equation}
\label{DiracMatrix}
C_{rr^\prime}={\{\varphi_r,\varphi_{r^\prime}\}}^*=
\left(\matrix{O& i\alpha \cr i\alpha & i(a^2-b^2)}\right),
\end{equation}
where $\alpha=-a\k\omega+bm$.

If now  we  consider  functions $ A^\prime, B^\prime$ which
  depend only on variables $x^i, {\cal P}_i, \xi^i$
 (e.g.  functions in (\ref{PrimedIndVar})),
 then taking  into    account   the  relations
\begin{equation}
{\{A^\prime, \xi_O\}}^*
={\{A^\prime, \xi_{D+1}\}}^* =0,
\end{equation}
and similar
relations   for $B^\prime$, it's easy   to    check  that
  the second  summand in  Eq. (\ref{FinalDiracBracket})
 consists of   only  one    term,     which contains the
 matrix   element $C^{-1}_{D+1,D+1}$.  Substituting   in  Eq.
 (\ref{FinalDiracBracket}) the expressions for $A^\prime$
 and $B^\prime$ from  (\ref{PrimedVar}) one   can    show
 by    direct calculations
that
\begin{equation}
\label{FinalDiracBrackets}
{\{A^\prime,B^\prime\}}_{D(\Phi)}=
{\{A,B\}}^*+
i{\{{\{A,B\}}^*,({\cal P}_j\xi_j)\}}^*
({\cal P}_j\xi_j)\frac{(b\k+a)}
 {\tilde \b(\omega + m)}.
\end{equation}
If now  we take  for
$A^\prime,B^\prime$ the variables $q, \pi , \psi$   from
 (\ref{PrimedIndVar}), then on account of
 (\ref{DiracBrackets}) we find from
 (\ref{FinalDiracBrackets})
\begin{eqnarray}
\label{CanonicalVar}
&&{\{\psi_i,\psi_j\}}_{D(\Phi)}=
{\{\xi_i,\xi_j\}}^*=-i\delta_{ij},\quad
{\{q_i,\pi_j\}}_{D(\Phi)}= {\{x_i,
{\cal P}_j\}}^*=-i\delta_{ij},\nonumber\\
&&{\{q_i,\psi_j\}}_{D(\Phi)}=
{\{x_i,\xi_j\}}^*=0,\quad {\{q_i,q_j\}}_{D(\Phi)}=
{\{x_i,x_j\}}^*=0,\\
&&{\{\pi_i,\pi_j\}}_{D(\Phi)}=gF_{ij}(x)-ig\pa _kF_{ij}\xi^k
({\cal P}_m\xi_m)\frac{(b\k+a)}
 {\tilde \b(\omega + m)}=gF_{ij}(q)\nonumber .
\end{eqnarray}

These relations  together with  (\ref{BracketProperty})
  prove that  the variables   $\tilde x_i,
  \tilde {\cal P}_j, \tilde \xi_k$
     are  Newton-Wigner  variables.
The  relations (\ref{CanonicalVar})  for  $D=4$ coincide
  with     similar relations obtained in  \cite{GG3} .
    Expressions of  initial variables   in terms of
  canonical   ones are given by the  following
\begin{eqnarray}
\label{InitialVar}
&& x_i=q_i-i\psi_i(\pi_j\psi_j) \frac{(a\k + b)}
{\tilde \a(\Omega + m)},  \nonumber\\
&&{\cal P}_i=\pi_i + igF_{ij}\psi_j
(\pi_k\psi_k)\frac{(a\k+b)}
{\tilde \a(\Omega + m)}, \\
&& \xi_i = \psi_i + \pi_i (\pi_j\psi_j)\frac{(a\k+b)}
{\tilde \a(\Omega+m)},  \nonumber
\end{eqnarray}
where  $\tilde \a=-a\k\Omega+bm, \Omega=\sqrt
{\pi_i^2+m^2+igF_{ij}\psi_i\psi_j}$.

Thus the variables $q_i, \pi_j , \psi_k$ have canonical
   Dirac brackets    and hence it is  convenient to quantize
the theory in  terms   of   these variables as  it was done
  for  $ D=4$ \cite{GG3}.
   Using (\ref{InitialVar})     and the
relations
\begin{equation}
 \xi_{D+1} =
-\frac{a(\pi_j\psi_j)}{\tilde \a},\quad
\xi_O=\frac{b(\pi_j\psi_j)}{\tilde \a},
\end{equation}
 one can  deduce the  expression for    the
  physical hamiltonian   of the spinning  particle in  the
 external   electromagnetic   field  in $D=2n$  in terms  of
canonical         variables
\begin{equation}
\label{Hamiltonian}
H_{\rm phys}=\Omega-g\k
A_O-ig\k\frac {F_{Ok}\psi_k(\pi_j\psi_j)}{\Omega
(\Omega+m)}.
\end{equation}
  It is   worth mentioning  that
  to quantize  the  theory   by  the   Berezin,  Marinov
prescription \cite{BM}  one must  expand $\Omega$, which enters
e.g.   (\ref{Hamiltonian} ),   in powers of
 $F_{ij}\psi_i\psi_j$, the  expansion terminating  in the
order $\frac{D-2}{2}$. If however  we are  interested in
the   terms in   the  hamiltonian which after  quantization
  will  be of  the order  of  $\hbar$, then  taking into
 account  that after  quantization
  $\psi_i \Rightarrow {\sqrt\frac{\hbar}{2}}\sigma_i$,
 we find    for   the quantum Hamiltonian the expression %
\begin{equation}
\hat H_{phys}{\rightarrow \atop \leftarrow}
\tilde \Omega -g\k A_O -ig\k\hbar
\frac{F_{Ok}\pi_j(\sigma_k\sigma_j-\sigma_j\sigma_k)}
{4\tilde \Omega(\tilde \Omega+m)} +
ig\hbar\frac{F_{ij}\sigma_i\sigma_j}{4\tilde \Omega},
\end{equation}
where $\tilde \Omega=\sqrt {\pi_i^2+m^2}$,
${\rightarrow \atop \leftarrow}$
 denotes the Weyl
correspondence between operators and their symbols.
Since the expressions for the initial
  variables in terms of canonical variables
 are form invariant in all dimensions, the equations of
 motion for $D=2n$ are similar to those
  for $D=4$ found in \cite{GG2}\cite{GG3}.

Thus the pseudoclassical Foldy-Wouthuysen
transformation significantly simplified
the search for the canonical variables and
enabled the quantization of $D=2n$ dimensional
 relativistic spinning particle in the external
 electromagnetic field .

\vspace{5mm}
\noindent{\em Acknowledgment}

The authors wish to thank I.V.Tyutin
for useful discussions. This research was
partly supported by the grant YPI-1993 of the
 "Bundesminister f\"ur Forschung und
Technologie", Federal Republic of
Germany.

\end{document}